\documentclass{article}
\usepackage{spconf,amsmath,graphicx}
\usepackage{url}
\usepackage{float}
\usepackage{siunitx}
\usepackage{amsfonts}
\usepackage{cite}

\title{LoopNet: Musical Loop Synthesis Conditioned On Intuitive Musical Parameters}
\name{Pritish Chandna\textsuperscript{1}, Ant\'onio Ramires\textsuperscript{1}, Xavier Serra \textsuperscript{1}, Emilia G\'omez\textsuperscript{1,2}\thanks{This work is partially supported by the Towards Richer Online Music Public-domain Archives (TROMPA) project and partly by the European Union’s Horizon 2020 research and innovation programme under the Marie Skłodowska-Curie grant agreement No765068, MIP-Frontiers. The TITANX used for this research was donated by the NVIDIA Corporation. }}
\address{\textsuperscript{1}Music Technology Group, Universitat Pompeu Fabra, Barcelona, Spain\\
\textsuperscript{2}Joint Research Centre, European Commission, Seville, Spain}

\begin{document}
\ninept

\maketitle
\begin{abstract}

Loops, seamlessly repeatable musical segments, are a cornerstone of modern music production. Contemporary artists often mix and match various sampled or pre-recorded loops based on musical criteria such as rhythm, harmony and timbral texture to create compositions. Taking such criteria into account, we present LoopNet, a feed-forward generative model for creating loops conditioned on intuitive parameters. We leverage Music Information Retrieval (MIR) models as well as a large collection of public loop samples in our study and use the Wave-U-Net architecture to map control parameters to audio. We also evaluate the quality of the generated audio and propose intuitive controls for composers to map the ideas in their minds to an audio loop. 

\end{abstract}
\begin{keywords}
Wave-U-Net, Loop Synthesis, Multi-Resolution Loss, Generative  Models,  Music Information Retrieval
\end{keywords}
\section{Introduction}
\label{sec:introduction}

Throughout human history, music has been an important cultural element in society. Fueled by technological advancements, the musical paradigm has evolved through the millennia, with changes in musical instruments for production and their arrangement in compositions. The barrier to entry for music creation has lowered significantly over the last few decades through the emergence of innovative electronic music production technologies including sampling and looping. 

The concept of loops is not new, repetitive units or motifs which are repeated through musical compositions have long been used in music across cultures, allowing composers to maintain continuity with various degrees of variability and complexity. This concept was further explored by artists like Pierre Schaeffer in the 1940s, who started repeating small audio segments within larger musical recordings to create compositions, leading to the emergence of loops. Loops can be defined as short audio excerpts which can be repeated from start to end in a seamless manner. They are often shared among composers and producers, through expanding community or commercial collections\footnote{Splice (\url{splice.com}) has over \num{300}k loops and Looperman (\url{looperman.com}) over \num{160}k.}.

In parallel, recent advancements in generative deep learning based methodologies have sparked a huge interest in music generation. Research has focused on both the symbolic~\cite{briot2017deep,boulanger2012modeling} and the audio domain~\cite{10.5555/3327757.3327895, Engel2020DDSP, carr2018generating, ramires2020neural, zukowski2018generating, engel2019gansynth, engel2017neural}. However, except for singing voice synthesizers, where the parameters of synthesis can be clearly defined in terms of the lyrics and score~\cite{chandna2019wgansing,BlaauwB2017_NeuralSynthesizer, BlaauwBD2019_DataEfficientVoiceCloning}, most of the proposed generative models for music rely on abstract latent embeddings for controlling the synthesis~\cite{engel2017neural, engel2019gansynth, Engel2020DDSP, carr2018generating, zukowski2018generating}. Efforts have been made to find semantic structures within the latent embeddings that can provide an intuitive control to the music producer, but there remains a gap between the intuition behind the parameters of the generative model and the perceptual qualities of the audio generated. To this end, some recent works have used Music Information Retrieval (MIR) techniques to generate single shot percussive sounds conditioned on semantically relevant musical features that are perceptually easier for the user to understand and manipulate~\cite{ramires2020neural, nistal2020drumgan}. In this paper, we aim to extend the generation of sounds conditioned on perceptually relevant features from one-shot sounds to loops. 

We model the waveform of a loop as a function of global features pertaining to timbre and harmony and time-varying features pertaining to rhythm. While there are several deep learning based architectural choices~\cite{10.5555/3327757.3327895, donahue2018adversarial, mehri2016samplernn} for modeling such a distribution, we decided to use a feed-forward convolutional design with skip and residual connections for the propagation of information between the layers of the network. This allows for the feed-forward generation of waveforms conditioned on the input. We leverage a curated collection of loop samples for training the model and evaluate our proposed methodology in terms of the perceived quality of the generated audio, the processing time required and the feature coherence between the input conditioning and the output sample. We also release our code publicly for reuse and provide interactive examples for a demonstration~\footnote{\url{github.com/aframires/drum-loop-synthesis}}.

The rest of the paper is structured as follows: Section \ref{sec:sota} presents some recent research done in similar fields followed by a brief overview of the dataset we use for training and testing our model in Section \ref{sec:datasets}. Our proposed methodology is outlined in Section \ref{sec:method} and a description of the evaluation strategy we use is presented in Section \ref{sec:evaluation}. This is followed by the results of our study in Section \ref{sec:results} and a discussion of the work presented here and how we plan to follow up in Section \ref{conclusion}.

\section{Related Works}
\label{sec:sota}
Audio synthesis technologies for music have been researched for many years, ranging from synthesizers generating pitched waveforms, to singing voice synthesizers conditioned on melody and text,\footnote{\url{vocaloid.com/en/}} to deep learning based models capable of generating entire songs~\cite{carr2018generating,zukowski2018generating}. For the context of this paper, we restrict ourselves to a description of generative models as deep learning based architectures used for musical audio synthesis. The most pertinent of such models are based on the autoregressive WaveNet~\cite{oord2016wavenet} and the SampleRNN~\cite{mehri2016samplernn} architectures. Initially proposed for generating high-quality speech samples, the WaveNet architecture models each sample of the speech waveform as a function of the previously predicted time steps, leading to the autoregressive nomenclature. The use of dilated causal convolutions allows the architecture to model longer term temporal dependencies between samples in an audio waveform than in the SampleRNN architecture. This architecture has subsequently been adapted for musical generation like singing voice synthesis conditioned on lyrics~\cite{BlaauwB2017_NeuralSynthesizer}, and instrument sound generation conditioned on the pitch and latent representations of timbre~\cite{engel2017neural, 10.5555/3327757.3327895, carr2018generating}. While the output of these models is subjectively similar to natural-sounding samples, the sequential nature of the model means that the processing time for generation is quite high, unless high-resource processing units are available. Feed-forward adaptations of the architecture via probability density distillation have been proposed showing that it is possible to directly map the input conditioning to the output waveform without sequentially predicting each sample of the waveform~\cite{oord2017parallel}. The WaveGAN~\cite{donahue2018adversarial} and GANSynth~\cite{engel2019gansynth} architectures are other examples of feed-forward networks capable of synthesizing an audio waveform. Both architectures are based on the adversarial training paradigm known as Generative Adversarial Networks (GANs)~\cite{goodfellow2014generative}. This methodology allows for minimization of the Jensen-Shannon divergence between the input data distribution and generated data distribution via a two network min-max scheme. 

Through these models, we can see that the most important factors for conditional waveform generation are the ability of the model to capture long-term temporal relationships between samples~\cite{10.5555/3327757.3327895} and to reproduce perceptually relevant features in the generated sample. One particular architecture capable of accounting for both these factors, particularly for a fixed-length input and output, in an efficient feed-forward manner, is the Wave-U-Net~\cite{stoller2018wave}. Originally proposed for waveform based source separation, the architecture consists of an encoder with a series of temporal convolutions, each capturing important information at a different temporal scale to produce a low-dimensional embedding. This embedding is then decoded via upsampling operations to the dimensions of the output waveform. Skip and residual connection are present between corresponding layers of the encoder and decoder to propagate information between the two stages of operation. Such architecture was recently used to produce high-quality single shot percussive sounds conditioned on timbral features in~\cite{ramires2020neural}. The authors used a waveform based loss augmented with a perceptually inspired loss taking into account the reproduction of spectral features of the output corresponding to the input. 

\section{Dataset Curation and Analysis}
\label{sec:datasets}
To collect a dataset for training the generative model, we use loops from an in-house collection of loops from Looperman\footnote{\url{looperman.com}}, a community loop database. This database contains loops provided by users with annotations for genre, instrumentation, key and tempo. 

We collected \num{8838} loops from the \textit{drums}\footnote{The instrumentation in the loops is not restricted to drums, even though the keyword is used for shortlisting. The loops might also include complementary melodic instruments on top of the drums.} category, with tempo annotations ranging from \num{120} to \num{140} beats per minute (BPM), the most frequent tempos in the collection. As tempo annotations provided by users might be noisy, we validate the tempo using a confidence measure~\cite{fontTempo}, taking into account the difference in the duration of a single bar and the relative duration of the entire loop. We only use loops which have a confidence measure higher than \num{99}\%, leading to a total of \num{8226} loops.

The Wave-U-Net architecture works with fixed-length input and output, requiring all loops to have the same length. To this end, we use the rubberband library\footnote{\url{breakfastquay.com/rubberband/}} to time-stretch all the loops to \num{130}BPM. We verified that this time-stretch did not create artifacts and that the loops were still sounding coherent after this step. To feed our model with fixed-length input, we split the loops into \num{1}-bar segments, with a duration of \SI{1.846}{\second} at \num{130}BPM. This process led to \num{49045} \num{1}-bar segments, which we then downsampled to \SI{16}{\kilo\hertz} and analyzed as shown in the next section. 

 We used \num{90}\% of the 1-bar loops for training and the rest for testing. As segments from the same loop are likely to be similar, we performed the training-test split before segmenting the audio into \num{1}-bar segments. As a result, there is no overlap between the loops from which segments were used for training and those which were used for testing.

\section{Proposed method}
\label{sec:method}
Once pre-processed, we analyze the loops to extract perceptually relevant features which we map back to the waveform via the neural network. We consider two types of features: local time-varying features, which are defined over the same temporal scale as the output and global features which can be summarized to be consistent across the length of the loop.

\subsection{Time-varying Conditioning Features}
\begin{figure}[h]
    \centering
    \includegraphics[width=0.48\textwidth]{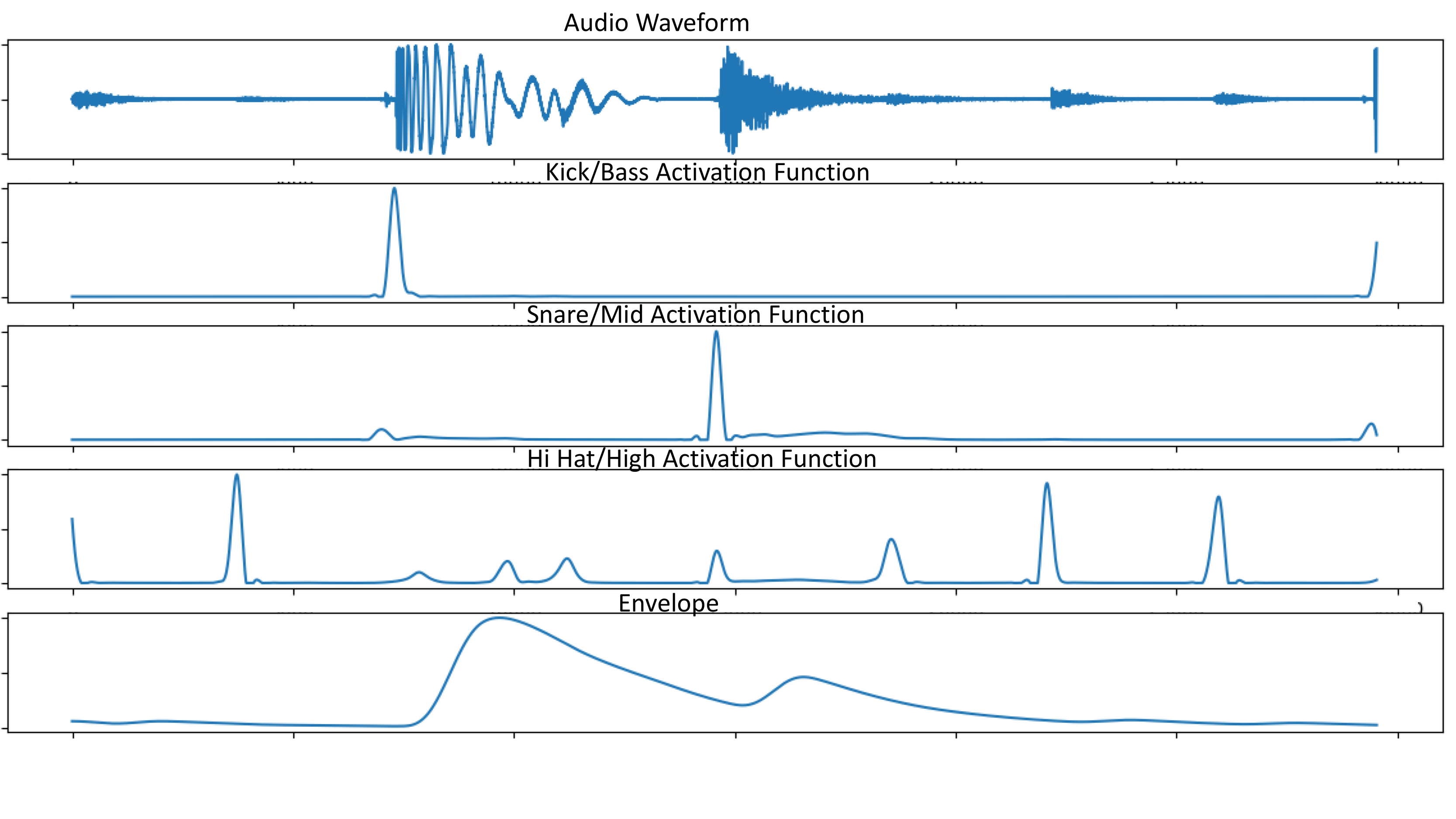}
    \caption{The waveform of the audio along with the activation function extracted using the Automatic Drum Transcription algorithm across the three bands and the energy envelope.}
    \label{fig:adt_output}
\end{figure}

Local time-varying features correspond to the rhythm of the loop. 
To obtain information on a loop's rhythmic pattern, we use an Automatic Drum Transcription algorithm~\cite{southall2017automatic}. This algorithm models the probability of a windowed frame of the analyzed audio having a kick drum, a snare drum or a hi-hat. We use one-dimensional spline interpolation \footnote{\url{docs.scipy.org/doc/scipy-0.18.1/reference/generated/scipy.interpolate.spline.html}} to interpolate the values from the windowed frames to the length of the waveform. An example of this representation for a loop in our dataset is presented in Figure \ref{fig:adt_output}. As seen in the figure, the activation function has sparse energy distribution across time, which does not represent the energy distribution of the loop. To account for this, we also calculated the envelope of the waveform, as used by~\cite{ramires2020neural}, with the same parameters. 

\subsection{Global Conditioning Features}\label{cond}
Global conditioning features correspond to textural features, both harmonic and non-harmonic, which can be assumed to be constant through the duration of the loop. 
To summarize the harmonic texture of the audio across the length of the loop, we used Harmonic Pitch Class Profiles (HPCP)~\cite{gomez2006tonal}\footnote{The Essentia implementation at \url{essentia.upf.edu/reference/streaming_HPCP.html} with the default parameters was used.}. This feature represents the energy distribution across the \num{12}-note chromatic scale commonly used in western music and is also intuitive for music makers. 
To model the abstract non-harmonic texture of the loop, we used the  perceptually pertinent timbral features proposed by Pearce et al.~\cite{pearce2017timbral,ramires2020neural}. These are hardness, depth, brightness, roughness, boominess, warmth and sharpness\footnote{To obtain these features we used an open-source implementation \url{github.com/AudioCommons/ac-audio-extractor}.}. We used the methodology proposed by~\cite{miron2013open}, to separately analyze the different frequency bands in each loop. The signal was analyzed through three different filters; a 1st order IIR low-pass filter with a cutoff frequency of \SI{90}{\hertz}, a 2nd order IIR band-pass centered in \SI{280}{\hertz}, and a 1st order IIR high-pass filter with cutoff frequency at \SI{9000}{\hertz}. The global conditioning feature for each of the frequency bands was summarized by taking an average across all windowed frames, concatenated, normalized and broadcast across time for conditioning the network as in~\cite{oord2016wavenet}. 

\subsection{Architecture}
\label{sec:architecture}
\begin{figure}[h]
    \centering
    \includegraphics[width=0.48\textwidth]{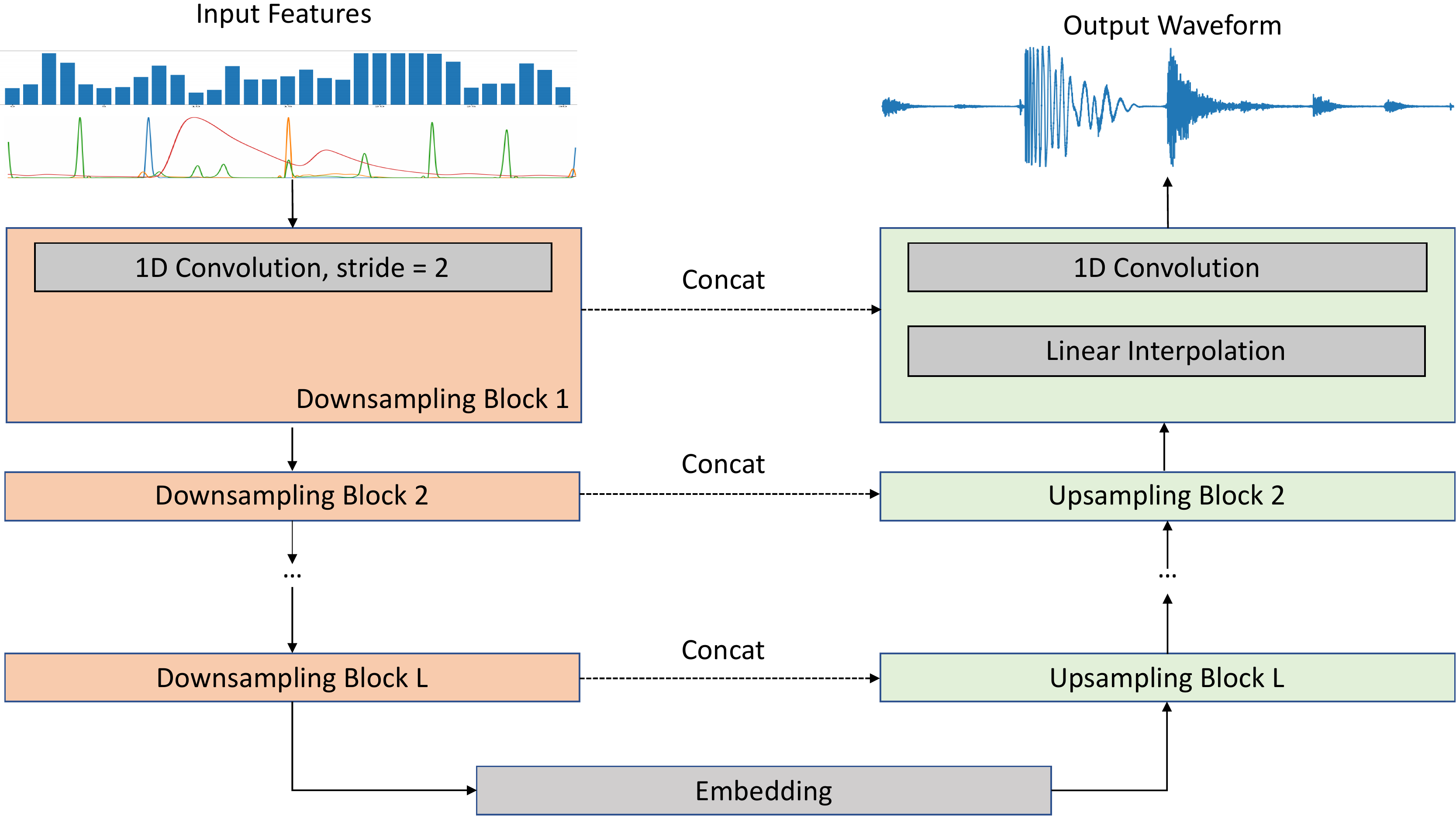}
    \caption{The Wave-U-Net architecture used in our study, the input includes the local conditioning with bandwise activation function and the global HPCP and timbral features which are broadcast along time. The output is the waveform of the loop.}
    \label{fig:archi}
\end{figure}

We use the feed-forward Wave-U-Net architecture~\cite{stoller2018wave}, used by~\cite{ramires2020neural}, for mapping the input features to the output waveform, $x$. The architecture is shown in Figure \ref{fig:archi}, and consists of an encoder and a decoder. The encoder downsamples the input via a series of convolutions with stride \num{2}, to produce a low-dimension embedding. A filter length of \num{5} is used throughout the layers and the number of filters is doubled after each \num{3} layers, starting with \num{32} filters. We use \num{10} layers in the encoder, to produce a low-dimensional embedding. This embedding is upsampled via linear interpolation~\cite{chandna2019wgansing,stoller2018wave} and each upsampling operation is followed by a $5\times1$ convolution to generate the output, $\hat{x}$. The inputs and outputs are vectors of length \num{29538}, representing the number of samples in \num{1} bar loops in our dataset. The corresponding layers in the encoder and decoder are connected via the concatenation of features to allow for the propagation of information~\cite{stoller2018wave}.

\subsection{Loss Functions} \label{sec:loss}
In this work, we experiment with several loss functions and compare their output quality. We used the reconstruction loss shown in Equation \ref{eq:loss1} and complemented it with a perceptually motivated loss based on the short time fourier transform (\textit{STFT}) based loss~\cite{ramires2020neural}, shown in in Equation \ref{eq:loss2}. We used a hopsize of \num{512} for calculating the STFT for $\mathcal{L}_{stft}$, over a FFT window resolution of \num{1024} samples, resulting in a frequency resolution of \SI{16.125}{\hertz} per bin. In addition, we used the multi-resolution loss~\cite{wangmulti, Engel2020DDSP}, which consists of calculating the STFT with various FFT window resolutions (\num{2048} samples for $i=0$, \num{1024} samples for $i=1$, \num{512} samples for $i=2$, \num{256} samples for $i=3$, \num{128} samples for $i=4$, \num{64} samples for $i=5$).

\begin{equation}
\mathcal{L}_{recon} = \mathbb{E} [\|\hat{x} - x\|_1] 
\label{eq:loss1}
\end{equation}
\begin{equation}
\mathcal{L}_{stft} = \mathbb{E} [\|\hat{x} - x\|_1] + \mathbb{E} [\|STFT(\hat{x}) - STFT(x)\|_1]  
\label{eq:loss2}
\end{equation}
\begin{equation}
\mathcal{L}_{multi} = \mathbb{E} [\|\hat{x} - x\|_1] + \sum\limits_{i=0}^5\mathbb{E} [\|STFT_i(\hat{x}) - STFT_i(x)\|_1] 
\label{eq:loss3}
\end{equation}

For the rest of the paper, we will refer to the model optimized using Equation \ref{eq:loss1} as \textit{WAV}, the model optimized for Equation \ref{eq:loss2} as \textit{WAVSPEC} and that optimized using the multi-resolution loss~\cite{Engel2020DDSP} shown in Equation \ref{eq:loss3} as \textit{MULTI}. 

\section{Experiments}
\label{sec:evaluation}
\subsection{Models}

In our experiment, we train and evaluate \num{5} different models. The first of the models, termed as \textit{STFT}, maps control features to the STFT of the waveform and uses the Griffin-Lim~\cite{griffin1984signal} algorithm for waveform reconstruction. The \textit{WAV}, \textit{WAVSPEC} and \textit{MULTI} models map the input features directly to the waveform and are optimized using the loss functions described in Section~\ref{sec:loss}. In addition, we train a model mapping the input features minus the envelope to the corresponding waveform, optimized using the multi-resolution loss. We term this model as \textit{MULTI NOENV}

\subsection{Parameters}
The network was trained using the Adam optimizer~\cite{kingma2014adam} with a batch size of \num{16}. Like~\cite{10.5555/3327757.3327895}, we found that after a certain limit of optimization, the loss function did not directly correspond to the audio quality of the output. As such, we heuristically selected the best epoch for synthesis for each of the models.

\subsection{Evaluation}

While there are several aspects of the generated output that can be assessed, we restrict our evaluation to two aspects: the audio quality and the coherence between changes in the input features to the output audio.

\subsubsection{Audio Quality Assessment}
Audio quality is a subjective attribute, which is difficult to quantitatively measure as each person listening to the audio is likely to have their own perception of the quality. Some quantitative metrics based on Deep Learning have been proposed recently to assign a quantitative degree to this attribute, including the \textit{Fr\'echet Audio Distance}~(FAD)~\cite{fad}, the \textit{Inception Score}~\cite{NIPS2016_6125} and the \textit{Kernel Inception Distance}(KID)~\cite{binkowski2018demystifying}. In this study, we use the FAD between the original test set and the generated output to assess the audio quality of the output. This metric can identify degradation of audio quality in a way related to human judgments, using audio-based latent embeddings from the Audioset~\cite{audioset} VGG'ish pre-trained model\footnote{\url{github.com/tensorflow/models/tree/master/research/audioset}}. To create a baseline for this evaluation, we resynthesized the audio using the Griffin-Lim~\cite{griffin1984signal} algorithm. This baseline is termed as \textit{Griffin-Lim} and has minimal degradation. 

\subsubsection{Timbral Feature Coherence}

We want changes in the input features to be reflected in the synthesis output. To this end, we assess feature coherence over \num{16} loops from the test set in a manner similar to~\cite{ramires2020neural,nistal2020drumgan}. We synthesized outputs changing each of the \num{21} timbral features individually, while keeping the rest unchanged. The feature was changed to three different values: \num{0.2}, \num{0.5} and \num{0.8}, over the normalized scale. We refer to this outputs as $\hat{x}_{low}^i$, $\hat{x}_{mid}^i$ and $\hat{x}_{high}^i$ and their corresponding features are $fs_{low}^i$, $fs_{mid}^i$ and $fs_{high}^i$ for the $i^{th}$ feature. This resulted in $16\times21\times3 = 1008$ different outputs for each model.
We then extracted timbral features as described in Section \ref{cond}.

We evaluate timbral feature coherence by validating if the models follow the order $fs_{high}^i >fs_{mid}^i>fs_{low}^i$. For this, we use three tests~\cite{ramires2020neural,nistal2020drumgan}: $E1$, which checks the condition $fs_{high}^i>fs_{low}^i$; $E2$, which checks $fs_{high}^i>fs_{mid}^i$; and $E3$, which checks $fs_{mid}^i>fs_{low}^i$ over all generated loops.

\section{Results and Discussion}
\label{sec:results}

\subsection{Audio Quality Assessment}
The results for the assessment of audio quality using the FAD metric are presented in Table \ref{table:FAD}. We encourage the reader to listen and subjectively evaluate the outputs for each model, provided in the accompanying website~\footnote{\url{github.com/aframires/drum-loop-synthesis}}. The website includes synthesis examples from the test set, as well as an illustration of user-interaction possibilities like timbre transfer

\begin{table}[H]
\centering
\begin{tabular}{l c  }

Model  &  FAD \\\hline
Griffin-Lim   & 1.26\\
STFT & 24.97\\
WAV & 13.83 \\
WAVSPEC & 9.06\\
MULTI   & 3.73 \\
MULTI NOENV    & 3.35 \\

\end{tabular}
\caption{FAD for the outputs of each of the models, when compared to the original test data. FAD values closer to 0 indicate higher similarity between the quality of the original audio and the assessed output.}
\label{table:FAD}
\end{table}

As seen in Table \ref{table:FAD}, the spectogram loss in the \emph{WAVSPEC} model leads to an improvement in synthesis quality over the \emph{WAV} and \emph{STFT} models. The use of the multi-resolution loss in the \emph{MULTI} models led to a significant improvement in audio quality measured by the FAD. We also see that the use of the overall envelope was redundant as the model was able to perform just as well when not conditioned on this feature. We stress that this measure is just a simplification of a highly subjective perceptual attribute and acknowledge that there is room for improvement in the audio quality. 

\subsection{Timbral Feature Coherence}
The results of the feature coherence evaluation are presented in Table \ref{table:coherence}. A high degree of coherence can be observed between changes in input features and the resulting output for all models except the \emph{WAV} model. The \emph{STFT} model outperforms the waveform based models for feature coherence but lags in audio quality.
\begin{table}[h]
\centering

\begin{tabular}{l c c c c }

&  & Accuracy &  \\ 
Model & E1 & E2 & E3 \\ \hline
STFT            & 87.50\% & 77.68\% & 79.46\% \\
WAV             & 67.44\% & 62.86\% & 52.14\% \\
WAVSPEC         & 76.79\% & 75.89\% & 74.11\% \\
MULTI           & 87.50\% & 77.68\% & 75.00\% \\
MULTI NOENV     & 83.93\% & 77.68\% & 75.00\% \\

\end{tabular}
\caption{Mean feature coherence across models for each error type.}
\label{table:coherence}
\end{table}

\section{Conclusion}\label{conclusion}
We present LoopNet, a deep learning based feed-forward loop synthesis algorithm, mapping intuitive input controls directly to the corresponding waveform. The nature of the model allows for fast synthesis of the loop, with processing time from control input to the synthesized output of the order of \SI{0.5}{\second} per loop, without the use of a GPU. In addition to the models presented in this study, we tried and tested various other loss functions, including training via the adversarial optimization and using DDSP~\cite{Engel2020DDSP} features for synthesis. In this study, we present some of the handpicked models which worked the best. We have verified that the models can maintain feature coherence between the control inputs and the generated output. 
The quality of the sound generated is a subjective aspect, that we have simplified and evaluated using a deep learning based methodology. The results are encouraging but there is room for improvement. 

The control features presented provide innovative avenues for user control over the synthesis. Apart from the straightforward mapping of MIDI input over a time grid to the input rhythm features, the ADT can also be computed directly from audio, allowing for mixing and matching of rhythm, timbre and harmonic features over loops or even over more abstract sounds such as human beatboxing or environmental sounds with interesting timbre. We provide examples of these on our complimentary website along with other interfaces for input control. We believe this will allow for a greater degree of freedom for musicians looking to expand their musical creativity while using loops for compositions. This also establishes a baseline for future works in loop synthesis looking to improve the quality of the loops synthesized. 

\pagebreak
\bibliographystyle{IEEEbib}
\bibliography{strings,refs}

\begin{thebibliography}{10}

\bibitem{briot2017deep}
Jean-Pierre Briot, et~al.,
\newblock Deep learning techniques for music generation--a survey,''
\newblock {\em arXiv preprint arXiv:1709.01620}, 2017.

\bibitem{boulanger2012modeling}
Nicolas Boulanger-Lewandowski, et~al.,
\newblock Modeling Temporal Dependencies in High-Dimensional Sequences:
  Application to Polyphonic Music Generation and Transcription,''
\newblock in {\em Proc. of the 29th International Conference on Machine
  Learning}, 2012, p. 1881–1888.

\bibitem{10.5555/3327757.3327895}
Sander Dieleman, et~al.,
\newblock The Challenge of Realistic Music Generation: Modelling Raw Audio at
  Scale,''
\newblock in {\em Proc. of the 32nd International Conference on Neural
  Information Processing Systems}, 2018, p. 8000–8010.

\bibitem{Engel2020DDSP}
Jesse~H. Engel, et~al.,
\newblock {DDSP:} Differentiable Digital Signal Processing,''
\newblock in {\em Proc. of the 8th International Conference on Learning
  Representations}. 2020.

\bibitem{carr2018generating}
CJ~Carr and Zack Zukowski,
\newblock Generating albums with SampleRNN to imitate metal, rock, and punk
  bands,''
\newblock {\em arXiv preprint arXiv:1811.06633}, 2018.

\bibitem{ramires2020neural}
Ant{\'o}nio Ramires, et~al.,
\newblock Neural Percussive Synthesis Parameterised by High-Level Timbral
  Features,''
\newblock in {\em Proc. of the 45th IEEE International Conference on Acoustics,
  Speech and Signal Processing}. 2020, pp. 786--790.

\bibitem{zukowski2018generating}
Zack Zukowski and CJ~Carr,
\newblock Generating black metal and math rock: Beyond bach, beethoven, and
  beatles,''
\newblock {\em arXiv preprint arXiv:1811.06639}, 2018.

\bibitem{engel2019gansynth}
Jesse~H. Engel, et~al.,
\newblock GANSynth: Adversarial Neural Audio Synthesis,''
\newblock in {\em Proc. of the 7th International Conference on Learning
  Representations}, 2019.

\bibitem{engel2017neural}
Jesse Engel, et~al.,
\newblock Neural audio synthesis of musical notes with wavenet autoencoders,''
\newblock in {\em Proc. of the 34th International Conference on Machine
  Learning-Volume 70}. 2017, pp. 1068--1077.

\bibitem{chandna2019wgansing}
Pritish Chandna, et~al.,
\newblock Wgansing: A multi-voice singing voice synthesizer based on the
  wasserstein-gan,''
\newblock in {\em Proc. of the 27th European Signal Processing Conference}.
  2019, pp. 1--5.

\bibitem{BlaauwB2017_NeuralSynthesizer}
Merlijn Blaauw and Jordi Bonada,
\newblock A Neural Parametric Singing Synthesizer Modeling Timbre and
  Expression from Natural Songs,''
\newblock {\em Applied Sciences}, vol. 7, no. 1313, 12/2017 2017.

\bibitem{BlaauwBD2019_DataEfficientVoiceCloning}
Merlijn Blaauw, et~al.,
\newblock Data Efficient Voice Cloning for Neural Singing Synthesis,''
\newblock in {\em {Proc. of the 44th IEEE International Conference on
  Acoustics, Speech, and Signal Processing }}, 2019.

\bibitem{nistal2020drumgan}
Javier Nistal, et~al.,
\newblock DrumGAN: Synthesis of drum sounds with timbral feature conditioning
  using Generative Adversarial Networks,''
\newblock in {\em Proc. of the 21st International Society for Music Information
  Retrieval Conference}, 2020.

\bibitem{donahue2018adversarial}
Chris Donahue, et~al.,
\newblock Adversarial Audio Synthesis,''
\newblock in {\em Proc. of the 7th International Conference on Learning
  Representations}. 2019.

\bibitem{mehri2016samplernn}
Soroush Mehri, et~al.,
\newblock SampleRNN: An Unconditional End-to-End Neural Audio Generation
  Model,''
\newblock in {\em Proc. of the 5th International Conference on Learning
  Representations}, 2017.

\bibitem{oord2016wavenet}
A{\"{a}}ron van~den Oord, et~al.,
\newblock WaveNet: {A} Generative Model for Raw Audio,''
\newblock in {\em Proc. of the 9th {ISCA} Speech Synthesis Workshop}, 2016, p.
  125.

\bibitem{oord2017parallel}
A{\"{a}}ron van~den Oord, et~al.,
\newblock Parallel WaveNet: Fast High-Fidelity Speech Synthesis,''
\newblock in {\em Proc. of the 35th International Conference on Machine
  Learning-Volume 80}, 2017, pp. 3915--3923.

\bibitem{goodfellow2014generative}
Ian Goodfellow, et~al.,
\newblock Generative adversarial nets,''
\newblock in {\em Advances in Neural Information Processing Systems}, 2014, pp.
  2672--2680.

\bibitem{stoller2018wave}
Daniel Stoller, et~al.,
\newblock Wave-U-Net: {A} Multi-Scale Neural Network for End-to-End Audio
  Source Separation,''
\newblock in {\em Proc. of the 19th International Society for Music Information
  Retrieval Conference}, 2018, pp. 334--340.

\bibitem{fontTempo}
Frederic Font and Xavier Serra,
\newblock Tempo Estimation for Music Loops and a Simple Confidence Measure,''
\newblock in {\em Proc. of the 17th International Society for Music Information
  Retrieval Conference}, 2016, pp. 269--275.

\bibitem{southall2017automatic}
Carl Southall, et~al.,
\newblock Automatic Drum Transcription for Polyphonic Recordings Using Soft
  Attention Mechanisms and Convolutional Neural Networks.,''
\newblock in {\em Proc. of the 18th International Society for Music Information
  Retrieval Conference}, 2017.

\bibitem{gomez2006tonal}
Emilia G{\'o}mez~Guti{\'e}rrez,
\newblock Tonal description of music audio signals,''
\newblock 2006.

\bibitem{pearce2017timbral}
Andy Pearce, et~al.,
\newblock Timbral Attributes for Sound Effect Library Searching,''
\newblock in {\em AES International Conference on Semantic Audio}, 2017.

\bibitem{miron2013open}
Marius Miron, et~al.,
\newblock An open-source drum transcription system for pure data and max msp,''
\newblock in {\em Proc. of the 38th IEEE International Conference on Acoustics,
  Speech and Signal Processing}. 2013, pp. 221--225.

\bibitem{wangmulti}
X.~{Wang}, et~al.,
\newblock Neural Source-Filter Waveform Models for Statistical Parametric
  Speech Synthesis,''
\newblock {\em IEEE/ACM Transactions on Audio, Speech, and Language
  Processing}, vol. 28, pp. 402--415, 2020.

\bibitem{griffin1984signal}
Daniel Griffin and Jae Lim,
\newblock Signal estimation from modified short-time Fourier transform,''
\newblock {\em IEEE Transactions on Acoustics, Speech, and Signal Processing},
  vol. 32, no. 2, pp. 236--243, 1984.

\bibitem{kingma2014adam}
Diederik~P. Kingma and Jimmy Ba,
\newblock Adam: {A} Method for Stochastic Optimization,''
\newblock in {\em Proc. of the 3rd International Conference on Learning
  Representations}, 2015.

\bibitem{fad}
Kevin Kilgour, et~al.,
\newblock Fr{\'{e}}chet Audio Distance: {A} Reference-Free Metric for
  Evaluating Music Enhancement Algorithms,''
\newblock in {\em 20th Annual Conference of the International Speech
  Communication Association}, 2019, pp. 2350--2354.

\bibitem{NIPS2016_6125}
Tim Salimans, et~al.,
\newblock Improved Techniques for Training GANs,''
\newblock in {\em Advances in Neural Information Processing Systems 29}, pp.
  2234--2242. 2016.

\bibitem{binkowski2018demystifying}
Mikołaj Bińkowski, et~al.,
\newblock Demystifying MMD GANs,''
\newblock in {\em Proc. of the 6th International Conference on Learning
  Representations}, 2018.

\bibitem{audioset}
J.~F. {Gemmeke}, et~al.,
\newblock Audio Set: An ontology and human-labeled dataset for audio events,''
\newblock in {\em Proc. of the 42nd IEEE International Conference on Acoustics,
  Speech and Signal Processing}, 2017, pp. 776--780.

\end{thebibliography}

\end{document}